\documentclass[preprint,showpacs,prb,floatfix,superscriptaddress,citeautoscript,cite]{revtex4}
\usepackage{graphicx}
\usepackage{color}
\usepackage{amsmath}

\begin{document}

\title{Single Layer PbI$_{2}$: Hydrogenation-driven Reconstructions}

\author{C. Bacaksiz}
\affiliation{Department of Physics, Izmir Institute of Technology, 35430 Izmir, Turkey}

\author{H. Sahin}
\affiliation{Department of Photonics, Izmir Institute of Technology, 35430 Izmir, Turkey}

\date{\today}


\begin{abstract}

By performing density functional theory-based calculations, we investigate how hydrogen atom interacts with the 
surfaces of monolayer PbI$_{2}$ and how one and two side hydrogenation modify its structural, electronic, and magnetic properties. Firstly, it was shown that T-phase of single layer PbI$_{2}$ is energetically favorable than the H-phase. It is found that hydrogenation of its surfaces is possible through the adsorption of each hydrogen on 
iodine sites. While H atoms do not form a particular bonding-type at low concentration, by increasing the number of hydrogenated I-sites well-ordered hydrogen patterns are formed on PbI$_{2}$ matrix. In addition,  we found that for one-side hydrogenation, the structure forms a (2$\times$1) Jahn-Teller type distorted structure and band gap is 
dramatically reduced compared to hydrogen-free single layer PbI$_{2}$. Moreover, in the case of full-hydrogenation, the  
structure also possesses another (2$\times$2) reconstruction with reduction in the band gap. 
Easy-tunable electronic and structural properties of single layer PbI$_{2}$ by hydrogenation reveal its potential use in 
nanoscale semiconducting device applications.
\end{abstract}

\maketitle

\section{Introduction}

After a decade of research which was triggered by the synthesis of graphene\cite{Novoselov2}, the theoretical prediction and synthesis of many  2D ultrathin materials  such as silicene,\cite{Cahangirov,Kara} germanene,\cite{Cahangirov,Davila,Yang} stanene,\cite{Guzman,Bechstedt} transition metal dichalcogenides (TMDs),\cite{Gordon,Coleman,can1,Ross,Sahin2,Tongay,Horzum,Chen3} hexagonal III-V binary compounds (h-BN, h-AlN)\cite{Sahin3,Kim,Tsipas,Bacaksiz} and metal hydroxides (Ca(OH)$_{2}$, Mg(OH)$_{2}$)\cite{Aierken,Suslu}  have been achieved. However, recent research efforts have been directed towards not only synthesis of graphene-like materials, but also towards the functionalization of existing ultra-thin crystal structures. These recent studies have revealed some important results such as (i) tunable bandgap opening in graphene\cite{Elias,Flores,Sahin5,Sahin6,Sofo,Sahin4} (ii) H-defect-induced magnetization of graphane,\cite{Sahin4,Zhou2} (iii) band gap engineering in silicene and germanene,\cite{Wang3,Houssa,Voon} (iv) stability enhancement in h-BN,\cite{Cabria} (v) tunable magnetic features in TMDs.\cite{Shi,Pan}

One of the most recently synthesized single layer semiconductor is lead iodide (PbI$_{2}$). In bulk PbI$_{2}$, which is 
a member of metal halides family, van der Waals stacked individual layers are in the form of octahedral T-phase. As a 
precursor material for lead iodide perovskites, given by the general formula of 
CH$_{3}$NH$_{3}$PbI$_{3-n}$X$_{n}$ (X=Cl, Br), 
PbI$_{2}$ has been used in many different device applications.\cite{Lee2,Jeng,Tan,Green,Guo} Recently, by performing the 
optical measurements and first-principles calculations of bulk and few-layer PbI$_{2}$, Toulouse \textit{et al.} showed 
that exciton binding energy dramatically increases with a decreasing number of layers.\cite{Toulouse}  In addition, the 
synthesis of the monolayer PbI$_{2}$ within the carbon nanotubes was reported by Cabana et al.\cite{Cabana} In a recent 
work, Zhou \textit{et al.} investigated the  structure, stability, electronic and optical properties of the monolayer 
PbI$_{2}$ and also the hetero-bilayer form with  graphene by using first principles calculations.\cite{Zhou} However, 
to our knowledge, the interaction between hydrogen (H) atom and the surface of monolayer PbI$_{2}$ and how the physical 
properties are effected under hydrogenation are still open questions. 

In this study, using first principles calculations based on density functional theory (DFT), we investigate the 
interaction of hydrogen atoms with bare single layer PbI$_{2}$. We also focus on structural and electronic properties of 
half- and fully-hydrogenated monolayers of PbI$_{2}$. We found that the T-phase  is energetically more favourable than 
H-phase and it is an indirect semiconductor. Our investigation revealed that the chemical functionalization by both 
half- and full-hydrogenation cause reconstruction in the structure of monolayer PbI$_{2}$ and lead to significant 
reduction in the band gap of the structure. Such a hydrogen-driven reconstruction in monolayers of PbI$_{2}$ have not 
been reported before.

The paper is organized as follows: in Sec. \ref{comp} we give details of our computational methodology. An overview of 
the structural phases and the electronic properties of monolayer hexagonal PbI$_{2}$ are presented in Sec. 
\ref{t-h}. In Sec. \ref{single} we focus on the interaction between the single hydrogen atom and the monolayer 
PbI$_{2}$. The effects of the one side hydrogen coverage, in other word half-hydrogenation, of monolayer PbI$_{2}$ are 
given in Sec.\ref{half}, and after that properties of fully-hydrogenated monolayer PbI$_{2}$ are presented in Sec. 
\ref{full}. Finally, we present our conclusion in Sec. \ref{conc}.

\section{Computational Methodology}\label{comp}

We perform the structural optimization and  molecular dynamics (MD) simulations by using the Vienna ab-initio 
simulation 
package, VASP\cite{vasp1,vasp2,vasp3} which is based on density functional theory (DFT). To describe electron exchange 
and correlation The Perdew-Burke-Ernzerhof (PBE) form of the generalized gradient approximation (GGA)\cite{GGA-PBE} was 
adopted. The vdW forces, which are effective on intralayer interaction, were taken into account by using the DFT-D2 
method of Grimme.\cite{vdW1,vdW2} To obtain the charge distribution of the configuration, a Bader charge analysis was 
used.\cite{Bader1,Bader2} The stability of the resulting structures was examined with ab-initio MD calculations.

The following parameters were used for analyses. The total energy difference between the sequential steps in the
iterations was taken 10$^{-5}$ units as convergence criterion. The convergence for the Hellmann-Feynman forces on each 
unitcell was taken to be 10$^{-4}$ eV/\AA{}. 0.05 eV of Gaussian smearing factor was used and the pressures 
on the unit cell were decreased to a value less then 1.0 kB in all three directions. For the determination of accurate 
charge densities, Brillouin zone integration was performed using a $12\times12\times1$ $\Gamma$-centered mesh for the 
primitive unit cell. To avoid interactions between adjacent PbI$_{2}$ configurations, our 
calculations were performed with a large unit cell including $∼$16 \AA{} vacuum space.

\begin{figure}
\includegraphics[width=8.5 cm]{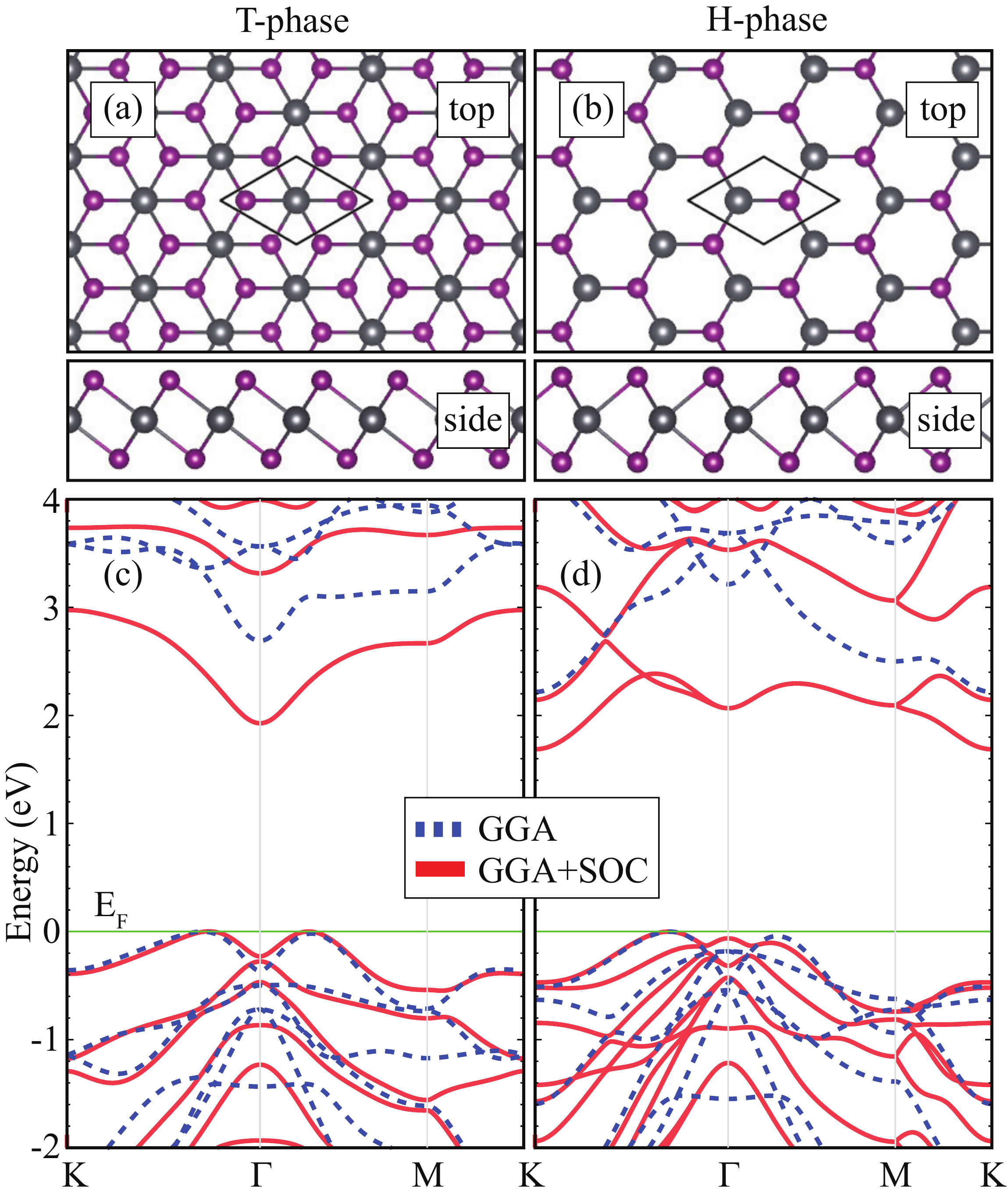}
\caption{\label{figt-h}
(Color online) (a) and (b) illustrate the structure of monolayer 1T and 1H 
PbI$_{2}$, respectively. (c) and (d) are the corresponding band structures where the blue and red curves 
represent the GGA and GGA+SOC, respectively. The Fermi level (E$_{F}$) is given by green solid line.}
\end{figure}

\begin{table}
\caption{\label{table1} Calculated parameters for monolayer PbI$_{2}$ are the lattice constant in the lateral 
direction, $a$; the atomic distance Pb and I atoms, d$_{Pb-I}$; the 
charge transfer from Pb to I atom, $\Delta\rho$; the work function $\Phi$; and the cohesive energy, E$_{c}$. 
E$_{g}^{\text{GGA}}$ and E$_{g}^{\text{GGA+SOC}}$ are the energy band gap values within GGA and GGA+SOC, respectively.}
\begin{tabular}{rcccccccc}
\hline\hline
 & a   & d$_{Pb-I}$ & $\Delta\rho$ &$\Phi$& E$_{c}$ & E$_{g}^{\text{GGA}}$ 
& 
E$_{g}^{\text{GGA+SOC}}$\\
 & (\AA) & (\AA) & ($e^{-}$) &(eV)& (eV) & (eV) & (eV) \\
\hline
1T PbI$_{2}$ & 4.44 & 3.23  & 0.9 &5.96& 2.76 & 2.69 & 1.93 \\
1H PbI$_{2}$ & 4.32 & 3.26  & 0.8 &6.05& 2.70 & 2.22 & 1.68 \\
\hline\hline
\end{tabular}
\end{table}

\section{H and T Phases of Single Layer PbI$_{2}$}\label{t-h}

Similar to TMDs, monolayer PbI$_{2}$ can form two different phases, octahedrally coordinated 1T and trigonal-prismatic 
coordinated 1H. As shown in Fig.~\ref{figt-h}, both phases have three trigonal subplanes where the Pb subplane is 
sandwiched by two I-subplanes. While the 1T phase is a member of the  $P\overline{3}m2$ space group where subplanes of 
it are $ABC$ stacked, the 1H is a member of the $P\overline{6}m2$ space group where subplanes of it are $ABA$ stacked. 
The lattice vectors of both phases are $\textbf{v}_{1}=a(\frac{1}{2},\frac{\sqrt{3}}{2},0)$, 
$\textbf{v}_{2}=a(\frac{1}{2},-\frac{\sqrt{3}}{2},0)$ where $|\textbf{v}_{1}|=|\textbf{v}_{2}|$ and $a$ is the lattice 
constant. The atomic coordinates of 1T phase are 
$(\frac{|v_{1}|}{2},\frac{|v_{1}|}{2},0)$, $(\frac{|v_{1}|}{6},\frac{|v_{1}|}{6},\frac{c}{2})$, and 
$(\frac{5|v_{1}|}{6},\frac{5|v_{1}|}{6},-\frac{c}{2})$ for the Pb and two I atoms, respectively, where $c$ is the 
distance between the subplanes of I atoms. The atomic coordinates of 1H phase are given as 
$(\frac{|v_{1}|}{3},\frac{|v_{1}|}{3},0)$, $(\frac{2|v_{1}|}{3},\frac{2|v_{1}|}{3},\frac{c}{2})$, and 
$(\frac{2|v_{1}|}{3},\frac{2|v_{1}|}{3},-\frac{c}{2})$. 

Lattice constants of the optimized crystal structures of 1T and 1H phases are 4.44 \AA{} and 4.32 \AA{}, respectively. 
However, as shown in Table \ref{table1}, interatomic distance d$_{Pb-I}$ of T-phase (3.23 \AA{}) is found to be 
shorter 
than H-phase (3.26 \AA{}). Total energy calculations also reveal that the 1T phase is $165$ meV per unit cell more 
favourable over the H-phase. The cohesive energies per atom of 1T and 1H phases are 2.76 and 2.70 eV, respectively. 
These results are consistent with the previous results which showed that the T-phase is the most favourable form of both 
bulk and monolayer crystals of PbI$_{2}$. In addition, the workfunctions ($\Phi$) of 1T and 1H phases are calculated to 
be 5.96 eV and 6.05 eV, respectively. 

We also present electronic energy band dispersions of 1T and 1H phases (approximated by GGA and GGA+SOC) in Figs. 
\ref{figt-h} (c) and (d), respectively. The  1T phase of PbI$_{2}$ monolayer has an indirect band gap where the valence 
band maximum (VBM) is between the $K$ and the $\Gamma$ symmetry points and the conduction band minimum (CBM) is at the 
$\Gamma$ point. As given in Table \ref{table1}, the energy bandgaps of 1T phase are 2.69 and 1.93 eV within GGA and 
GGA+SOC. Total charge donated by each Pb atom are 0.9 and 0.8 $e^{-}$ for 1T and 1H phases, respectively. Moreover, the 
1H phase has also an indirect band gap where the VBM is between the $K$ and the $\Gamma$ points and CBM is at the $K$ 
point. The band gap of 1H phase is calculated to be 2.22 eV within GGA and 1.68 eV within GGA+SOC.

It is seen that although the T and H phases have similar electronic characteristics at the valence band edges which are 
immeasurable by experimental tools such as ARPES, lattice parameter and the work function is unique feature of each 
phases. Discussions in the following chapters will be performed on the 1T-phase that corresponds to the ground state 
crystalline structure of single layer PbI$_{2}$.

 \section{Interaction with Single Hydrogen} \label{single}
 
For engineering the structural, electronic  and magnetic properties of a material, surface hydrogenation is an easy and 
powerful method. From theoretical point of view determination of the interaction between the PbI$_{2}$ surface and H 
atoms is of importance.

\begin{figure}
\includegraphics[width=8.5 cm]{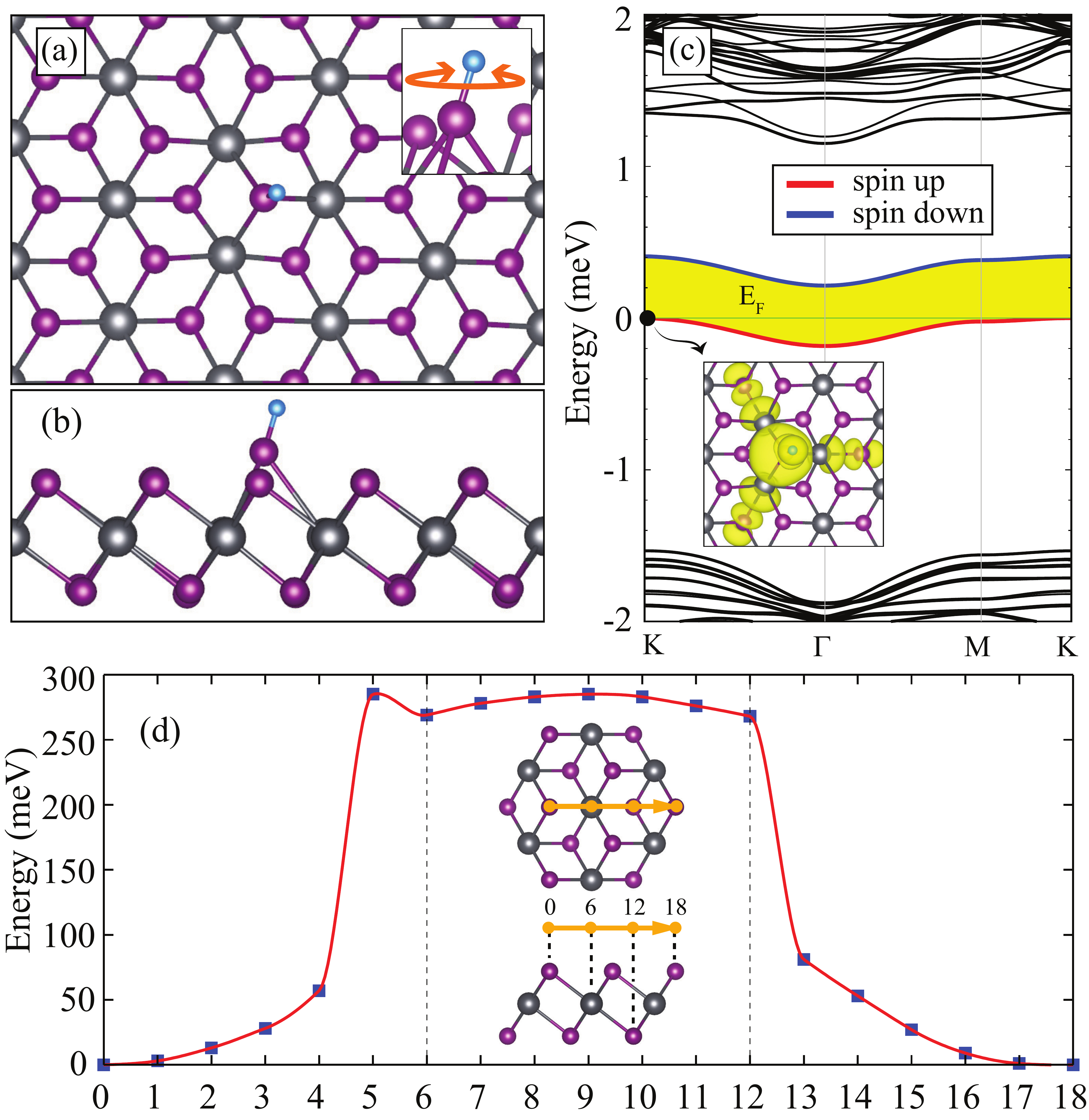}
\caption{\label{figsingle}
(Color online) (a) and (b) 
show the structure of H atom adsorbed ($3\times3$) PbI$_{2}$ from top and the side, respectively. The inset in 
(a) illustrates the rotation of H around the I atom.  (c) is the band structure of single H atom adsorbed $3\times3$ 
PbI$_{2}$ monolayer where the blue and 
red curves represent the spin-different states. Inset shows the charge distribution of VBM at the $\Gamma$. (d) is the 
diffusion barrier plot through the given path. The Fermi level (E$_{F}$) is given by green solid line.}
\end{figure}

For the calculation of adsorption and diffusion characteristics of H atom on the surface of monolayer PbI$_{2}$ a 
3$\times$3 supercell, which is enough to avoid the interaction between adjacent H atoms, is used.  First of all, to 
determine the most favourable adsorption site of H atom, various initial positions over the surface are calculated; 
top-Pb site, midpoint of Pb-I bond, the top-I site and the sites in between these points. As shown in Fig. 
\ref{figsingle} (a), adsorption of hydrogen atom in the vicinity of I atom with the bond length of 1.66 \AA{} is much 
more preferable than adsorption on other lattice points. It is also seen that the formation of tilted I-H bond with 
the surface leads to slightly out-of-plane relaxation of the underlying I atom. Considering its spin polarized ground 
state in vacuum, the binding energy of single H atom is calculated to be 0.32 eV which is quite small compared to 
binding energy of H to graphene (0.98 eV\cite{Sahin7}).

In addition, our Bader charge analysis reveal that bonded H atom preserves its 1.0 $e^{-}$, on the other hand, the H 
bonded I atom has 7.0 $e^{-}$ in contrast with the other I atoms which have 7.4 $e^{-}$. The displaced 0.4 $e^{-}$ from 
that particular I atom are shared by the nearest three Pb atoms. Our analysis reveals that the H atom do not form a 
bond when the vdW term excluded. Therefore, one can conclude the bonding type of H with PbI$_{2}$ surface as a weak vdW 
type bonding.

The single H bonded system have midgap states which originate from H 
atom. When we consider these states, the band gap is 0.21 eV as shown in Fig \ref{figsingle} (c). 
There is one localized band near the Fermi level which have charge distribution as shown inset of Fig \ref{figsingle} 
(c).  The system has a magnetic ground state with magnetic moment of 1.0 $\mu_{B}$. It also appears that only the 
states around Fermi level have splitting due to hydrogen-induced magnetism.

To have more general picture of interaction between H atom and the monolayer PbI$_{2}$, we perform the diffusion 
barrier calculation which is shown in Fig. \ref{figsingle} (d). From the top of an I atom to top of its second nearest I 
atom, the energy difference plot shows that top of I atom possesses the minimum energy. The barrier to escape from the 
influence of I atom is around 285 meV. There are one local minimum on the path which coincide the center of triangle 
of I atoms over the Pb atom. In addition, considering the energy plot around the I atom as an harmonic potential, the 
jump frequency of H atom is estimated to be $\nu\approx0.191$ GHz. These results are consistent 
with the MD 
simulations reveal that at low temperatures (up to 50 K), each single H rotates almost freely around the I atom. 
Although each H is bonded to underlying I, there is no certain preferable bonding direction with the PbI$_{2}$ surface. 

 \section{One-side Hydrogenation} \label{half}
Following the analysis of the interaction of H atoms with the surfaces of PbI$_{2}$, in this section, we investigate 
how the structural, electronic and magnetic properties are modified upon the hydrogen coverage of surfaces. Firstly, 
the 
half-hydrogenated (half-H-PbI$_{2}$),  one-by-one hydrogen coverage of each I atom on the same surface, structure is 
investigated. 

\begin{figure}
\includegraphics[width=8.5 cm]{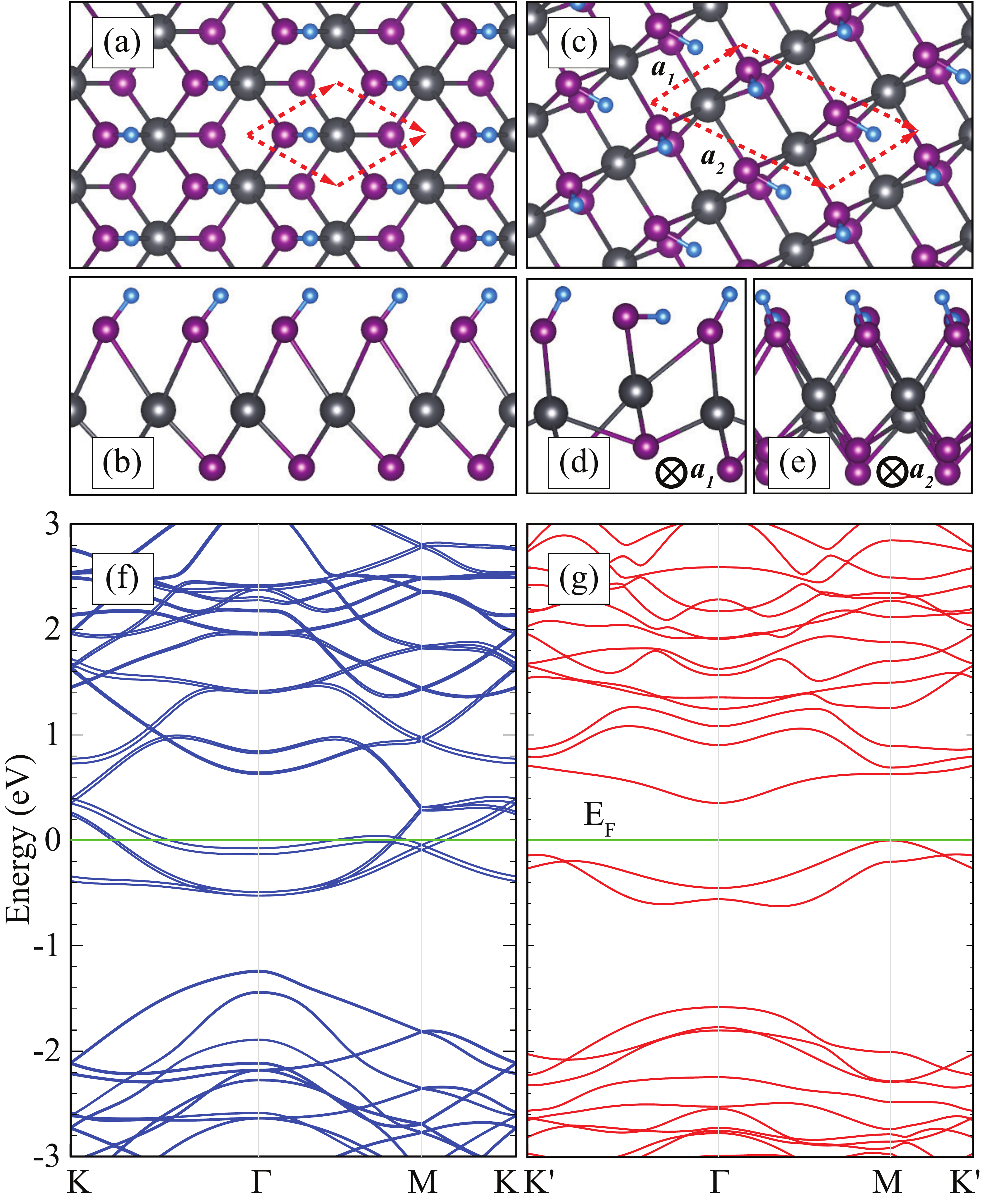}
\caption{\label{fighalf}
(Color online) The structures in (a) and (b) for perfect hexagonal and (c), (d), and (e) for distorted forms of  
half-hydrogenated monolayer PbI$_{2}$ (half-H-PbI$_{2}$), respectively. View direction are given in (d) and (e). (f) 
and (g) are the corresponding band structures. To compare the perfect and distorted forms, 2$\times$2 
unitcells are used for both band structures. The Fermi level (E$_{F}$) is given by green solid line. }
\end{figure}

As shown in Figs. \ref{fighalf} (a) and (b), H atoms form a well-ordered crystal structure (where 
$\theta$=60$^{\circ}$) 
in the one-side covered structure. The lattice constant of this perfectly hexagonal crystal structure is calculated to 
be 4.05 \AA{}. Compared to hydrogen-free bare PbI$_{2}$ structure, Pb-I bond length 
increases from 3.23 to  3.77 \AA{} at the hydrogenated side. However, at the hydrogen-free side, Pb-I bonds become 
stronger and the bond length is calculated to be 3.08 \AA{}. Moreover, alongside with the structural changes, the 
half-hydrogenation also modifies charge distribution in the crystal structure. Bader charge analysis shows that 
functionalization cancels charge transfer between Pb and I atoms at the hydrogenated side and therefore, removes the 
ionic character of Pb-I bond. In contrast, at the bare side of the half-H-PbI$_{2}$, the charge sharing 
increases where the I and Pb atoms have 7.4 and 3.5 $e^{-}$, respectively. Electronically, as shown in Fig. 
\ref{fighalf} (f), the semiconducting character vanishes after hydrogenation and the system turns into a ferromagnetic 
metal with 0.14 $\mu_{B}$ per unitcell. However, hydrogenation-induced metalization in single layer PbI$_{2}$ requires 
further analysis.
 
\begin{table}
\caption{\label{table2} Calculated parameters for half- and full-hydrogenated monolayer PbI$_{2}$ (half-H-PbI$_{2}$ and 
full-H-PbI$_{2}$) are; the lattice constant in the lateral directions, \textbf{a}$_{1}$ and \textbf{a}$_{2}$; the angle 
between the lattice vectors, $\theta$; the work function $\Phi$; the cohesive energy, E$_{c}$; the binding energy per H 
atom, E$_{bind}$; the band gap, E$_{g}^{\text{GGA}}$; and magnetic moment per primitive cell, $\mu$.}
\begin{tabular}{lcccccccc}
\hline\hline
  & \textbf{a}$_{1}$& \textbf{a}$_{2}$& $\theta$  & $\Phi$ & E$_{c}$ &E$_{bind}$ & E$_{g}^{\text{GGA}}$&$\mu$\\
  & (\AA)           & (\AA)           &($^{\circ}$)&  (eV) & (eV)    &   (eV)      & (eV)        & ($\mu_{B}$) \\
\hline
half-H-PbI$_{2}$  & 4.22   & 7.83      & 58.9      & 3.57   & 2.23    & 0.64        &    0.35       &  0.0 \\
full-H-PbI$_{2}$  & 8.29   & 8.02     & 63.0       & 3.73   & 2.03    & 0.93       &    0.76        & 0.0 \\
\hline\hline
\end{tabular}
\end{table}

For further analysis of the half-hydrogenated structure we also perform finite temperature MD 
calculations. Our calculations showed that starting from very low temperatures (20-25 K) half-H-PbI$_{2}$ 
tends to undergo a structural transformation. Apparently, the perfectly hexagonal half-hydrogenated structure mentioned 
before corresponds to a local minimum on Born-Oppenheimer surface and for the determination of the ground state 
structure one needs small perturbation such as temperature. Then, by performing full structural optimization of the 
distorted structure created by MD calculation, we obtained the real ground state crystal structure of half-H-PbI$_{2}$.

As shown in Figs. \ref{fighalf} (c) and (d), the ground state of half-H-PbI$_{2}$ forms a (2$\times$1) reconstructed 
surface. It is calculated that this (2$\times$1) reconstructed phase of half-H-PbI$_{2}$ is 0.41 eV per unitcell more 
favourable than the perfectly hexagonal phase which is obtained by a total energy calculation performed at 0 K. In 
this half-H-PbI$_{2}$ structure, there are two bonding types of H atom. One H atom stands over the I atom with an angle 
of 28.5$^{\circ}$ with the normal of the structure plane, the other H reclines parallel to the surface at level of the 
I atoms.  As given in Table \ref{table2}, the lattice constants are 4.22 and 7.83 \AA{} with an angle of 58.9$^{\circ}$ 
and the workfunction at the hydrogenated side is found to be 3.57 eV. On the other side, the H binding energy per H 
atom is 0.64 eV which is much larger than that of single H on 3x3 supercell (E$_{bind}$=0.32 eV for single H). 

The charge distributions of the reconstructed structure are specific for the structurally different Pb, I, and H 
atoms. In general, with respect to H atom bonding configurations, all atoms have specific charge in (2$\times$1) 
reconstracted 
unitcell (see Fig. \ref{fighalf} (c), (d), and (e)). Shortly, the standing- and reclining-H atoms have 1.1 and 1.0 
$e^{-}$; the I atoms bonded with standing- and reclining-H have 7.0 and 7.1 $e^{-}$; the Pb atoms close to standing- 
and reclining-H have 3.4 and 3.7 $e^{-}$, respectively. On the bare side of the half-H-PbI$_{2}$, the I atoms have 
larger charge of 7.3 and 7.4 $e^{-}$ which are aligned in \textit{z}-axis with standing- and reclining-H, respectively

Furthermore, upon the temperature-driven structural transformation from perfectly hexagonal to (2$\times$1) 
reconstructed phase, not only the structure but also electronic properties are modified dramatically. As shown in Fig. 
\ref{fighalf} (g), the reconstruction removes the metallic property and the system becomes semiconductor with a small 
band gap of 0.35 eV. The band gap is still indirect in which the VBM and CBM are at M and $\Gamma$ points, 
respectively. After the reconstruction the system has a nonmagnetic ground state. This type of distortions are known 
as 
Jahn-Teller distortion in which the dangling bonds are filled up and the system transforms from metal to semiconductor. 
Therefore, semiconducting nature of (2$\times$1) reconstructed  phase of half-H-PbI$_{2}$, that corresponds to the 
ground state structure, is quite important for nanoscale optoelectronic device applications.

\section{Full Hydrogenation}\label{full}

As well as one-side hydrogenation that may be realized on one-side-supported material, one can also achieve both-side 
coverage of monolayer PbI$_{2}$ (full-H-PbI$_{2}$) when it is in freestanding form. Structural analysis of the 
full-H-PbI$_{2}$ is performed by applying the same methodology. MD calculations revealed that the ground state atomic 
structure of the  full-H-PbI$_{2}$ is quite different from  half-H-PbI$_{2}$.

\begin{figure}
\includegraphics[width=8.5 cm]{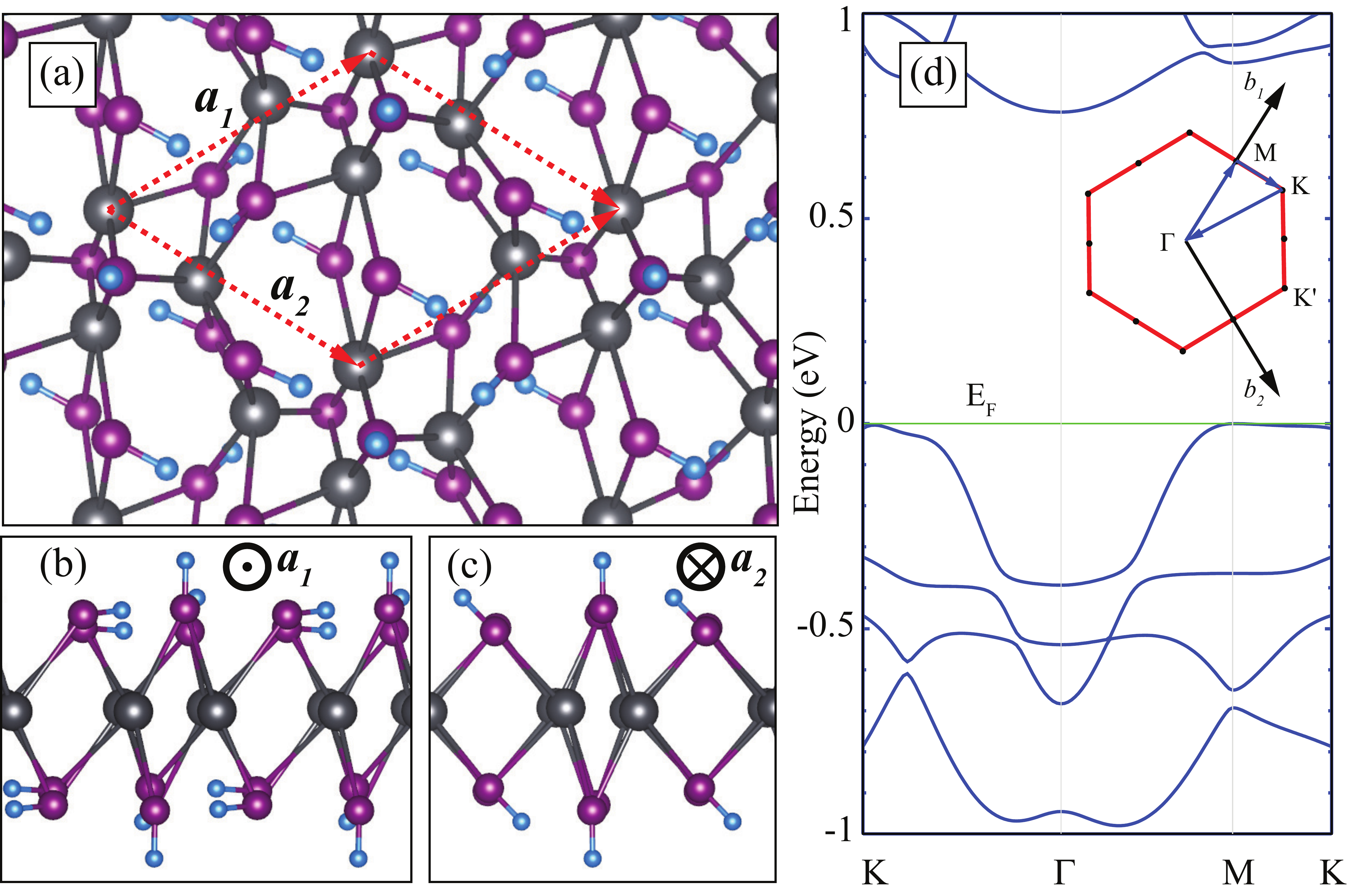}
\caption{\label{figfull}
(Color online) (a), (b), and (c) illustrate the structure of full-hydrogenated monolayer 
PbI$_{2}$ from top, side views with the corresponding directions, respectively. (d) is the 
band structure of full-hydrogenated monolayer PbI$_{2}$. The Fermi level (E$_{F}$) is given by green solid line.}
\end{figure}

Similar to half-hydrogenated structure, H-induced reconstructions take place starting from 20-25 K and therefore three 
different types of I-H bonds (perpendicular, tilted bond with 44.8$^{\circ}$, and the reclined parallel to the surface) 
are formed over the PbI$_{2}$ surface (see Figs. \ref{figfull} (a), (b) and (c)). Then the analysis of fully-optimized 
crystal structure reveals that the primitive unitcell of full-H-PbI$_{2}$ contains a (2$\times$2) reconstructed phase of 
PbI$_{2}$H$_{2}$. As given in the Table \ref{table2}, lattice parameters of the (2$\times$2) reconstructed structure are 
calculated to be 8.29 and 8.02 \AA{} with an angle of 63.0$^{\circ}$.

Upon full hydrogenation, not only the structure but also work functions, cohesive energies and the H binding energies 
significantly differ from half-hydrogenated structure. The work function of the full-H-PbI$_{2}$ is found to be 3.73 
eV. 
The cohesive energy of the full-H-PbI$_{2}$ is calculated to be 2.03 eV smaller than that of half-H-PbI$_{2}$. 
The calculated average binding energy of an H atom is 0.93 eV which is larger as compared to those of single H and 
half-hydrogenated cases. Here, it is noteworthy that the larger binding energy is consistent with the slightly larger 
work function. As shown in Figs. \ref{figfull} (d) the full-H-PbI$_{2}$ is a semiconductor with an indirect band gap of 
0.76 eV. 
The VBM and CBM stay at M and $\Gamma$ points which are similar to those of half-H-PbI$_{2}$. It is also seen that the 
Jahn-Teller type distortion in the structure removes dangling bonds and therefore the system is converged to 
non-magnetic semiconducting ground state.
ed

\section{Conclusion}\label{conc}

In this paper, we studied the newly emerging ultra-thin PbI$_{2}$ which is a member of metal halides. Starting from 
the comparison of two possible monolayer phases, 1T and 1H, interaction with single H atom, half- and 
full-hydrogenation of monolayer PbI$_{2}$ were investigated using first principles DFT calculations. The 1T phase was 
found to be favourable with the band gap of 2.69 and 1.93 eV within GGA and GGA+SOC, respectively. Our calculations 
also 
showed that the H atoms strongly reacts with the PbI$_{2}$ surface and further functionalization of its surfaces can be 
realized. The binding energy was found to be 0.3 eV.  

In the case of half-hydrogenation, it was found that the (2$\times$1) Jahn-Teller type distorted ground state structure 
corresponds to the ground state phase. We showed that reconstruction from perfectly hexagonal structure to 2$\times$1 
half-H-PbI$_{2}$ leads to dramatical modifications such as (i) metal-to-semiconductor transition (ii) removal of 
magnetic moments . Moreover, our calculations revealed that full-hydrogenated monolayer PbI$_{2}$ forms another 
(2$\times$2)  Jahn-Teller distorted form in its ground state. In that case, the structure, full-H-PbI$_{2}$, was a 
nonmagnetic semiconductor with a bandgap of 0.76 eV. Our results reveal that hydrogenation is an efficient way to 
engineer the structural and electronic properties of single layer PbI$_{2}$.

\section{Acknowledgments} 

This work was supported by the bilateral project between TUBITAK (through Grant  No. 113T050) and the Flemish Science 
Foundation (FWO-Vl). The calculations were performed at TUBITAK ULAKBIM, High Performance and Grid Computing Center  
(TR-Grid e-Infrastructure). CB and HS acknowledge the support from TUBITAK Project No 114F397. H.S. acknowledges support 
from Bilim Akademisi - The Science Academy, Turkey under the BAGEP program.

\end{document}